# Persistent Cell Motion in the Absence of External Signals: A Search Strategy for Eukaryotic Cells

Liang Li[1], Simon F. Nørrelykke[2], and Edward C. Cox[2*]

1 Department of Physics, Princeton University, Princeton, New Jersey, USA,
2 Department of Molecular Biology, Princeton University, Princeton, New Jersey, USA

Eukaryotic cells are large enough to detect signals and then orient to them by differentiating the signal strength across the length and breadth of the cell. Amoebae, fibroblasts, neutrophils and growth cones all behave in this way. Little is known however about cell motion and searching behavior in the absence of a signal. Is individual cell motion best characterized as a random walk? Do individual cells have a search strategy when they are beyond the range of the signal they would otherwise move toward? Here we ask if single, isolated, *Dictyostelium* and *Polysphondylium* amoebae bias their motion in the absence of external cues. We placed single well-isolated *Dictyostelium* and *Polysphondylium* cells on a nutrient-free agar surface and followed them at 10 sec intervals for ~10 hr, then analyzed their motion with respect to velocity, turning angle, persistence length, and persistence time, comparing the results to the expectation for a variety of different types of random motion. We find that amoeboid behavior is well described by a special kind of random motion: Amoebae show a long persistence time (~10 min) beyond which they start to lose their direction; they move forward in a zig-zag manner; and they make turns every 1-2 min on average. They bias their motion by remembering the last turn and turning away from it. Interpreting the motion as consisting of runs and turns, the duration of a run and the amplitude of a turn are both found to be exponentially distributed. We show that this behavior greatly improves their chances of finding a target relative to performing a random walk. We believe that other eukaryotic cells may employ a strategy similar to *Dictyostelium* when seeking conditions or signal sources not yet within range of their detection system.

## INTRODUCTION

It is generally believed that eukaryotic cells are large enough to detect and then move toward a signal by counting receptor occupancy. This can work because the relatively large eukaryotic cell is not subject to Brownian motion and can therefore use spatial differentiation to detect the direction of the signal over the relevant time scale. How this spatial differentiating is accomplished is an active area of research in fibroblasts, neutrophils and *Dictyostelium*, where the major components of the chemotactic response system are well known.

Of comparable interest is the behavior of these cells in the *absence* of a chemotactic (or other) signal. We might imagine, for example, that cells move about randomly in such a situation (Fig. 1A), or that they have evolved a strategy that somehow optimizes their chances of finding the source of the signal, even when they cannot sense it (Fig. 1B-D).

Is there indeed such a thing as an optimum search strategy? Recent theoretical work has suggested that a Lévy walk is the optimum for revisitable targets, that is, targets that repopulate at the same location after a period of time [1-4]. A Lévy walk is a special class of random walks whose step lengths ($l$) are best described by a power-law: $N(l) \sim l^{-a}$ where $2 < a < 3$. Thus there is no intrinsic scale to the step lengths, and very long steps can occur (Fig. 1B). Although there was thought to be experimental evidence for Lévy walk behavior in animal populations, a recent reanalysis of the data makes this unlikely [5], but see also [6]. In a search for non-replenishable targets, where, like hide and seek, each target can be found only once, it has been suggested that a two-state model optimizes the search [7-9] (Fig. 1C). A searcher alternates between a local random search and a fast linear relocation. Target detection does not occur during the linear phase, both phases stop at random times, and each new phase is initiated in a random direction. It has been suggested that such intermittent behavior may be used by foraging animals [10,11].

Do any of these processes describe the behavior of single cells searching for a hidden target? A great deal is known about how neurons [12], amoebae [13], and fibroblasts [14] find their targets once the signal has been sensed. In all three cases, more or less linear trajectories with variable low amplitude random behavior is the likely rule once the target is in range. But before pioneer neurons sense and begin to move up (or away from) a graded signal, do they send out filopodia at random, or do they bias their search to enhance the chances of finding the as yet undetected target? Do neutrophils wander at random before they detect bacterial peptides, or do they bias their motion

* Email: ecox@princeton.edu

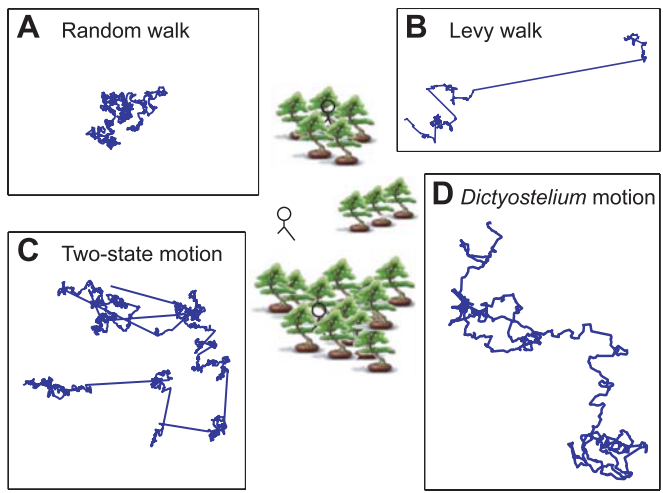

**Figure 1. The search problem and search models.** Four characteristic types of random motion: (**A**) Random walk. (**B**) Levy walk. Step lengths were picked from a power-law distribution, and thus very long steps are possible. A Levy walk is considered to be the best strategy for searching revisitable scarce targets. (**C**) Two-state motion. Here a ballistic relocate phase is followed by a diffusive search phase. Switching between states occurs at random times and in random directions. This model is believed to optimize the search for low density non-revisitable targets, for example, hide and seek in the patchy environment shown here. (**D**) Simulation of *Dictyostelium* searching based on features reported in this study. The speed and number of steps is the same in A and B.

in some fashion that provides a more efficient search algorithm?

We studied these questions by placing well-separated *Dictyostelium* amoebae on an agar surface free of food at a density of ~1 cell/cm$^2$, ~1000 cell diameters between cells, a distance chosen so that the amoebae could not sense and signal to each other. Our results are of two kinds: First, these cells show long directional persistence. They bias their motion by making turns every 1-2 min, remembering their last turn and turning away from it in a zigzag fashion. Similar results were obtained with the distantly related slime mold *Polysphondylium*. We provide a model that satisfactorily captures the turning bias of freely moving cells, and links short and long-term cell motion persistence times. Second, although we cannot say that this behavior has been optimized by selection, we do demonstrate that it is only somewhat less efficient than straight-line behavior.

We believe this is the first experimental evidence for a biased walk in a foraging eukaryotic cell in the absence of spatial and temporal cues. Because the machinery underlying eukaryotic cell motion has been so highly conserved during evolution, we think it is likely that similar behavior is characteristic of other target-seeking eukaryotic cells.

## RESULTS
### Nomenclature
Depending on which branch of science the reader hails from, the meaning of terms such as "random walk" and "Brownian motion" may differ. To eliminate at least this source of possible confusion we offer our own definitions here.

*Brownian motion* refers only to the passive random motion, reported by Robert Brown, of particles suspended in a fluid.

*Random walk* is taken to mean a stochastic path consisting of a series of steps, whose direction is chosen at random and where all directions are equally probable. The step size can be either random or fixed.

*Random motion* is the most general term and refers to any stochastic path describing the motion of a particle. There may or may not be a preferred spatial direction, correlations in step size, persistence in direction of motion, oscillations in the velocity, *etc*. The only demand is that there be some element of stochasticity in the motion.

### A long directional persistence
In Fig. 2A we show the behavior of 3 representative *Dictyostelium* cells, each one followed for 8-10 hrs corresponding to ~300 cell lengths, with a sampling interval of 10 sec. We found that amoebae traveled at an average speed of 7 μm/min for up to 10 hr (Fig. 2B), demonstrating that they maintained an adequate energy source over the course of these experiments. Thus, our modeling is not confounded by changes in average cell speed over time. On the time scale of minutes the speed was found to fluctuate around this average speed (Fig. 2B, insert).

The mean-squared displacements (MSD) of the individual cells are summarized in Fig. 3A and B as $\langle \delta(\tau)^2 \rangle$ vs $\tau$, and $\langle \delta(\tau)^2 \rangle / \tau$ vs $\tau$ respectively. $\tau$ is the time interval between any two positions, $\delta(\tau) = |\vec{r}(t+\tau) - \vec{r}(t)|$, and $\delta(\tau)^2$ was averaged over all pairs of time points for each trajectory. For a random walk, *e.g.* Brownian motion, $\langle \delta(\tau)^2 \rangle / \tau$ is constant. The 3 cyan curves are measurements from the trajectories shown in Fig. 2A, and the red curves are from an additional 9 trajectories. For $\tau < 30$ min, cell movement deviates significantly from the random walk expectation and is essentially ballistic, *i.e.*, the cells are on average moving in a straight line with constant velocity. For time intervals $\tau$ between 10 and 100 min, the population-averaged data were well fitted by an exponential cross-over from directed motion to a random walk, $\langle \delta(\tau)^2 \rangle = 2t_p v^2 (\tau - t_p(1-\exp(-\tau/t_p)))$ [15], where $v = 5.4 \pm 0.1$ μm/min is a characteristic speed and $t_p = 8.8 \pm 0.1$ min is a persistence time (Fig. 3B, yellow curve). Note that a cell displaces itself approximately 3 full cell diameters (~50 μm) in 8.8 min. This transition from directed to random walk is characteristic of the entire time record of more than 10 hr, and the persistent time is independent of where on the trajectory we begin our analysis.

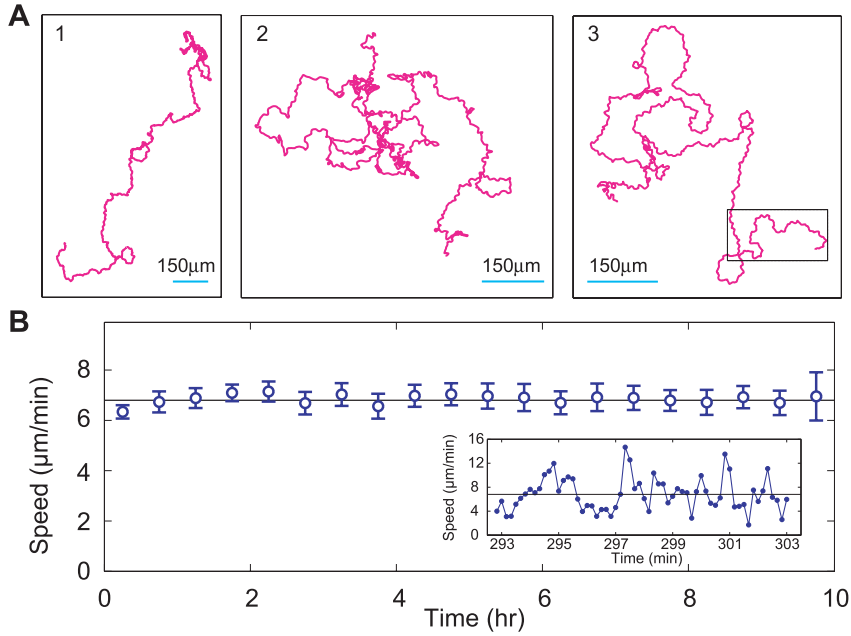

**Figure 2. Cell trajectories and speeds.** (**A**) Three typical 10hr cell trajectories. Boxed regime, see Fig. 5 caption. (**B**) Cells do not slow down over the ten-hour observation time, so we can think of them as being in a stationary (time-independent) state. However, on the time-scale of minutes the speeds do show fluctuations around their average, time-independent values (see insert). The error bars were obtained by first using a 30 min window to average each of twelve trajectories, and then, for each 30 min average, calculating the standard error of the twelve averages.

We calculated cell velocities as $\vec{v}(t) = (\vec{r}(t) - \vec{r}(t-\tau))/\tau$ for different values of $\tau$ and plotted $v_x$ vs $v_y$ (Fig. 4). With increasing $\tau$, bigger and bigger gaps appear in the centers of these plots. At very much larger values of $\tau$, greater than 30 min, the distribution of $v_x$ vs $v_y$ values again approaches Gaussian behavior, as expected (Fig. 4). As we show in the discussion, these results essentially rule out two well-understood models of random motion, worm-like-chain (WLC) [15] and Ornstein-Uhlenbeck (OU) [16] models for *Dictyostelium* cell trajectories.

### Angular changes and cell motion

In order to quantify the behavior of a cell we first introduce a measure of the cell's instantaneous direction of motion (Fig. 5). This measure is chosen as the *cumulative* angle $\varphi$, between the cell's velocity vector and a fixed direction in space: If, initially $\varphi(t=0) = \varphi_0$, and the cell at some time $t$ later has moved through a complete, counter-clockwise, circle, then the new direction of motion is $\varphi(t) = \varphi_0 + 2\pi$. That is, the angle $\varphi$ tracks not only the instantaneous direction of cell motion, but also the winding number of the cell trajectory, and thus, to an extent, the history of the cell's directional changes.

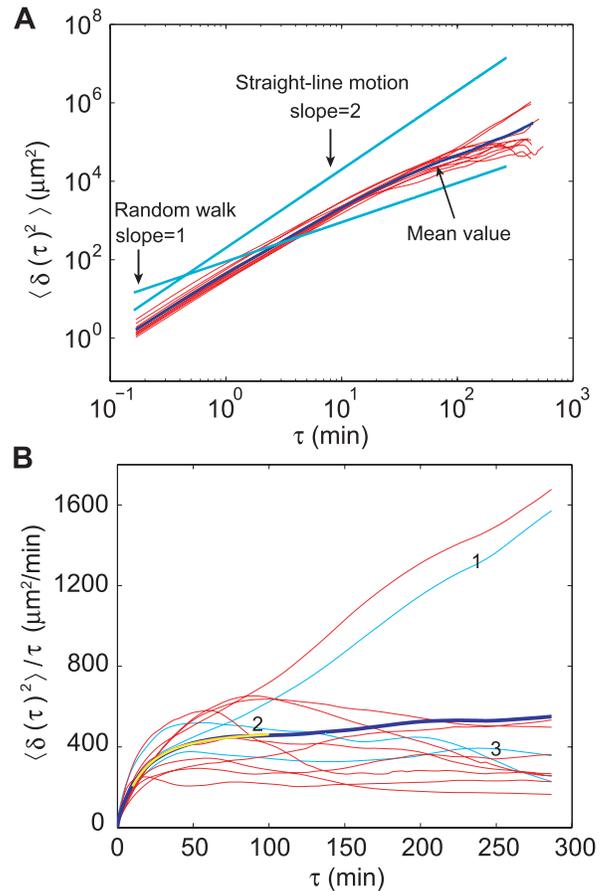

**Figure 3. Mean-squared displacement.** (**A**) Log-log plot of the mean-squared displacement *vs* time interval $\tau$. (**B**) Mean-squared displacement divided by $\tau$ plotted as a function of $\tau$. Random walk would gives rise to a line with zero slope. Cyan, data from the 3 trajectories showed in Fig. 2; Red, additional 9 trajectories; Blue, average of all 12 trajectories. Yellow, fit of an exponential cross-over from directed to random walk in the interval $\tau$ [10:100] min: $\langle \delta(\tau)^2 \rangle = 2t_p v^2(\tau - t_p(1-\exp(-\tau/t_p)))$, where $v = 5.4 \pm 0.1$ μm/min is a characteristic speed, and $t_p = 8.8 \pm 0.1$ min is a persistence time.

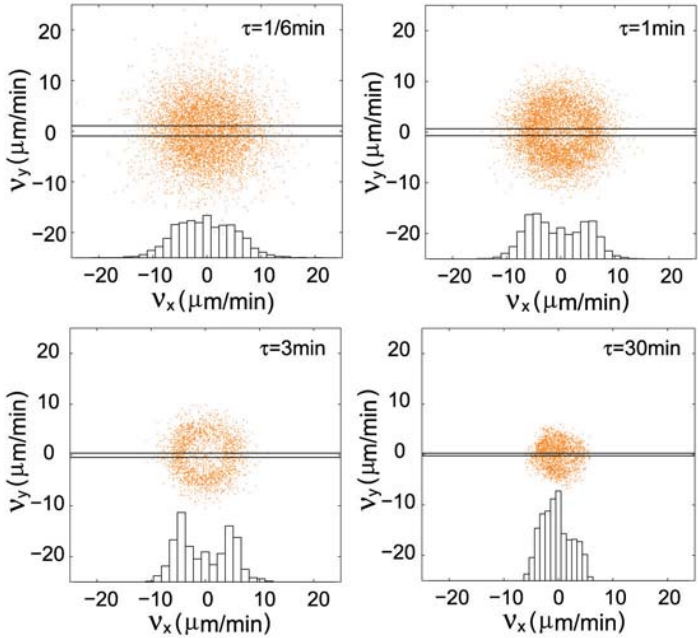

**Figure 4. Non-Gaussian velocity distribution.** Velocities were calculated for different $\tau$: $\vec{v}(t) = (\vec{r}_t - \vec{r}_{t-\tau})/\tau$ and $v_x$ was plotted vs $v_y$ with increasing $\tau$. Larger and larger gaps at the centers of the distributions with time demonstrate that the cell velocity distributions are non-Gaussian. As expected, at very large $\tau$, the distribution approaches a Gaussian again. Inserts, histograms of the $x$ component of the velocities for the intervals defined by the parallel lines.

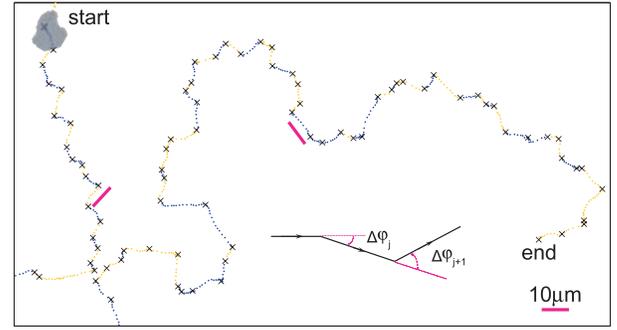

**Figure 5. Cells move in a zig-zag manner.** Enlarged view of the rectangular box in Fig. 2. The magenta scale-bars are 10μm. Turns are marked by black crosses. Motion following left turns is blue and motion following right turns is yellow.

With this definition, periods of straight-line motion correspond to a *constant* angle; small zig-zags in the direction of motion show up as *oscillations* in the angle (Fig. 6C); large turns as large *increases* and *decreases* in the angle (red box, Fig. 6A); and circling behavior adds or subtracts $2\pi$ per circle completed in the counter-clockwise and clockwise direction (Fig. 6B), respectively. In Fig. 6 we see all of these classes of behavior except the first one: The cells are never observed to move in a completely straight line. However, any given cell will at times appear to move in a certain fixed direction $\theta$, around which the instantaneous angle $\varphi$ fluctuates. In Fig. 6C, the cell maintained an average direction $\theta = -\pi/4$ for about an hour and $\varphi$ fluctuated around this value with a characteristic time of 2-3 min and amplitude near $\pi/4$.

**Turning preference**

To further investigate cell motion, we studied the occurrence of discrete turns in the cell trajectories. Over a range of user-defined parameters, such as the threshold for calling a turn, the results reported below are robust (see Materials and Methods).

We observed the same frequency of left and right turns, and thus choosing one hand over the other does not contribute to the observed wiggling behavior. However, these data do reveal a strong turning preference, in which cells tend to turn away from their last turn. Fig. 7A illustrates our turn-run-turn analysis. Fig. 7B plots data from all 12 experimental runs. The turning ratio was biased by a factor of $2.1 \pm 0.1$ (mean ± sem, n = 12 cells), obtained by classifying 4822 consecutive turns from all trajectories. The correlation coefficient for consecutive turning directions, for all cells, was -0.36, with a P-value $<10^{-4}$, a highly significant anti-correlation. There is a weak, but significant, positive correlation between the last and the second-from-last turn, but cell memory does not extend much further back (Fig. 7C). For comparison, the insert in Fig. 7C shows the autocorrelation from a Monte Carlo simulation of the WLC model.

Fig. 7D shows a histogram of the turn amplitudes. Its tail is best described by an exponential distribution (characteristic angle $\approx 0.67$ rad). Small angles are rarely observed, partly because of limitations in the turn detection algorithm, but also because of the pseudopod-branching process we discuss later. No significant temporal correlation was observed (upper right panel). Fig. 7E is the histogram of time intervals between turns. It is also an exponential distribution (characteristic time $\approx 0.67$ min) and again no significant autocorrelation was observed, consistent with a Poisson process. Fig. 7F gives the histogram of distances between the positions of the cells at consecutive turns. Its tail is well fitted

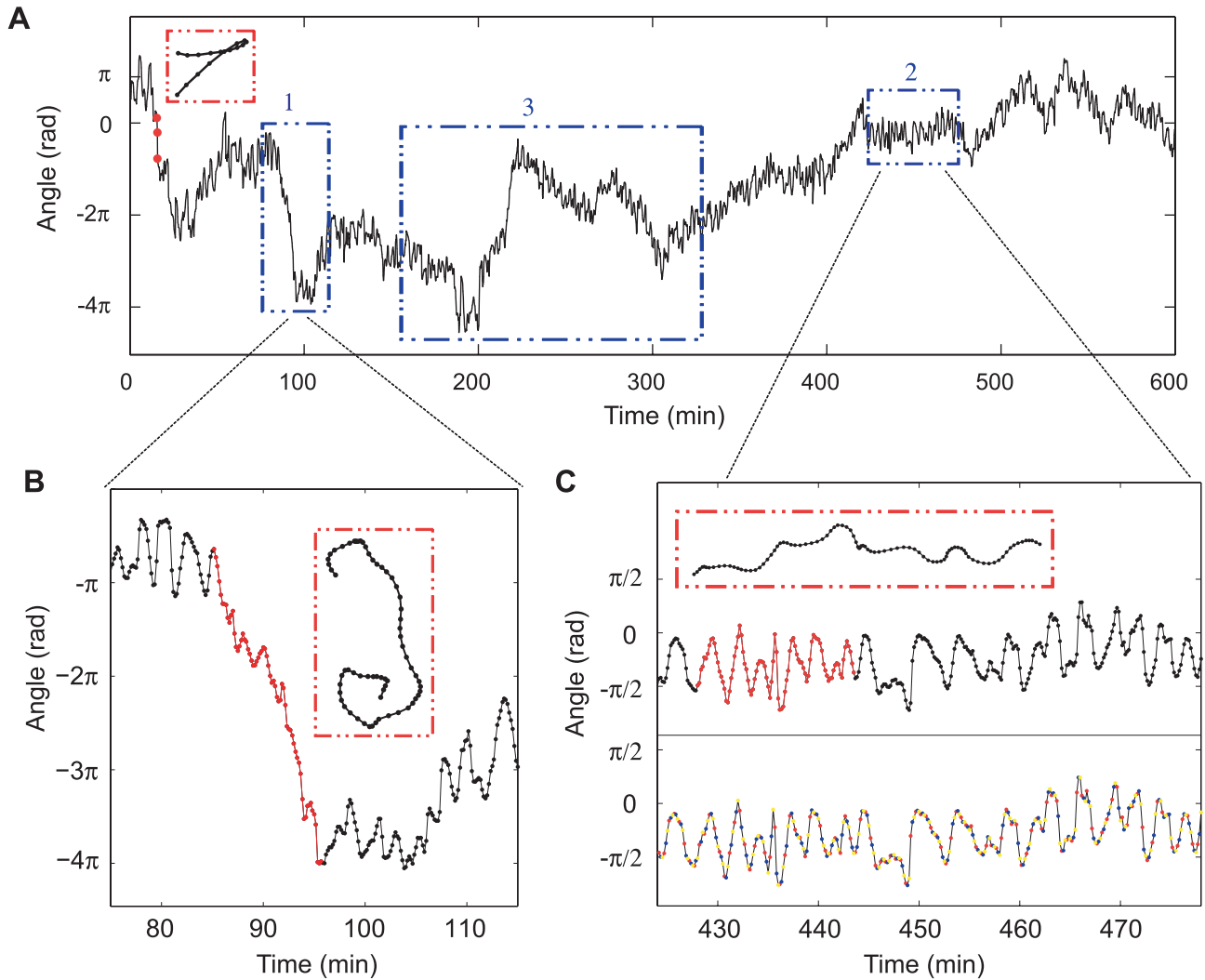

**Figure 6. Cumulative angles. (A)** Angles were corrected by +/- 2π for changes larger than π. As shown by trajectories in red boxes, sharp turns correspond to large drops or rises; continuous turns in one direction appear as continuous drops or rises (box 1, enlarged view in B); and periods of directed motion are summarized as plateaus (box 2, enlarged view in C). On average, angles change but slowly with time. Box 3, see Fig. 8 caption. **(B)** Enlarged view of box 1 in A. The transition from 0 to -4π is continuous. **(C)** Enlarged view of box 2 in A. To reduce the influence of noise, angles were calculated at a larger τ (30s) for most of the analyses. Thus from each trajectory, 3 interlaced time-series of angles were obtained. They are marked by different colors in the lower panel.

by an exponential distribution (characteristic length ≈ 5 μm).

**Biased motion in *Polysphondylium***
These experiments were repeated with the distantly related slime mold amoeba *Polysphodylium pallidum* with essentially the same results (Figure S1, S2). *Polysphodylium* and *Dictyostelium* are in different Genera, use different chemotactic signals, have different fruiting body morphologies, and diverged ~500,000 years ago [17]. This suggests that the dynamic behavior documented here is highly conserved and suggests further that it may be a feature of all migratory eukaryotic cells.

## DISCUSSION
**Non Ornstein-Uhlenbeck (OU) and non worm-like-chain (WLC) motion**
We have learned that *Dictyostelium* cell motion cannot be described by a random walk. Next, we show that two other standard models also fail to capture our data. First, note that an exponential crossover of the form we used to fit the MSD data (Fig. 3B) arises in several models describing disparate physical phenomena: Both WLC [15] models from polymer physics, and OU [16] processes from the modeling of Brownian motion, have this feature. Thus, an exponential crossover is not in itself enough to pin down the dynamics of cell motion.

We observed larger and larger gaps appearing in the centers of $v_x$ vs $v_y$ plots, with increasing $\tau$ (Fig. 4). This is because at very short time intervals the cells have moved very little, and the data shows essentially random micron-scale jiggling about the centroid of the cell. This Gaussian behavior at very small $\tau$ also indicates a non-WLC movement, because a WLC motion has a constant speed, thus giving rise to a circular plot. With increasing $\tau$, the behavior is distinctly non-Gaussian, ruling out the OU process because it predicts a Gaussian distribution for all $\tau$. Two-dimensional velocity histograms with a crater shape are one of the hallmarks of self-propelled persistent motion and have been predicted to occur when cells migrate [18]. To the best of our knowledge, this is the first published experimental demonstration of this effect.

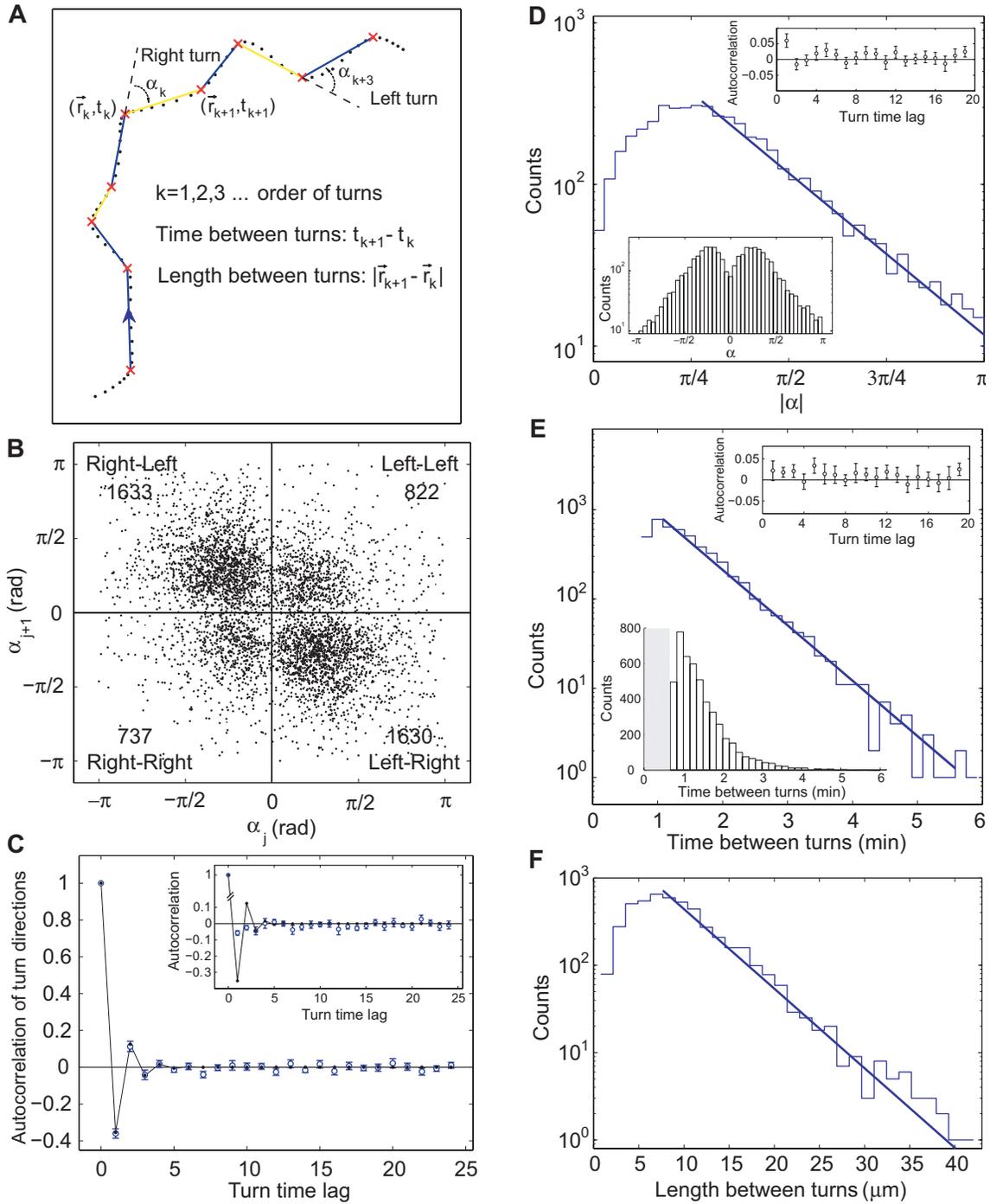

**Figure 7. A left turn is followed by a right turn – a Poisson process.** (**A**) Definition of angle of turns ($\alpha$), direction of turns (left or right), time between turns and length between turns. (**B**) The $j^{th}$ turn plotted against the $(j+1)^{th}$ turn for the data from all 12 trajectories. There are 3263 data points in the second and fourth quadrant, 1559 in the first and third, and thus the $(j+1)^{th}$ turn is biased by the $j^{th}$ turn by a factor of 2.1. (**C**) Autocorrelation function for the turn directions (see text for details). Blue: Experimental values and standard errors. Black: Theoretical expectation value for a Markov process with probabilities taken from panel B (see text for details). Insert: Verification that turn-correlations are real and not an artifact of the turn-detection algorithm. Blue: Autocorrelation function for synthetic data. The angle-dynamics was simulated by a worm-like-chain model (WLC) with parameters taken from the MSD of the real data. A small, negative, artifactual correlation is detected which extends for around 3 turns. Black: Same as the main-panel, shown for comparison. (**D**) Histogram of turn amplitudes. Its tail is well fitted by an exponential distribution (characteristic angle = 0.67 rad). The rounding off at small values is caused by thresholding in the turn-detection algorithm and this sharp cut-off is smoothed by the coarse-graining applied when calculating the angles. Lower left panel: Histogram of $\alpha$. Upper right panel: Autocorrelation function for turn amplitudes, no correlation was observed. The positive value at time-lag one is a verified artifact of the turn-detection algorithm. (**E**) Histogram of time intervals between detected turns. These data are well fitted by an exponential distribution (characteristic time = 0.67 min). Data is from all 12 trajectories. Lower left panel: Same histogram but on linear scale. The smallest detected value for $t_{j+1}-t_j$ is 40 sec, the cut-off shown by the grey bar. Upper right panel: Normalized autocorrelation function for time between turns. No significant correlations were observed, consistent with a Poisson process. (**F**) Histogram of length between turns. Its tail is well fitted by an exponential distribution (characteristic length = 5 μm). Distribution of length is trivially exponential if cell averaged speed is constant and times between turns are exponentially distributed.

**Modeling angular change**

To capture the behavior observed in cumulative angles (Fig. 6), we write down dynamic equations for the angles $\varphi$ and $\theta$. The average direction of motion $\theta$ is modeled as a random walk, reflecting the experimental fact that there are no preferred directions of motion in the absence of a chemotactic signal. The instantaneous direction of motion $\varphi$, was observed to be somewhat enslaved by $\theta$, exhibiting noisy oscillations around it with a characteristic time-scale. We therefore model $\varphi$ as a sum of a random walk $\theta$, and noisy oscillations $y$ (Eqs. 1-3).

Fig. 8 illustrates the biological processes and features corresponding to the two terms in Eqs. 1 and 2. In our model, each cell has an "intrinsic vector" that establishes an angle $\theta$ relative to an arbitrary, but fixed, direction in space. We leave the molecular components of this intrinsic vector undefined, but remark that it could *e.g.* be given by a vector connecting the cell nucleus with the centrosome [19]. This intrinsic vector is assumed to change its orientation only slowly over time, and it consequently does not track the instantaneous direction of motion, which will fluctuate around $\theta$. The instantaneous direction of motion $\varphi$, is determined by short-lived processes, such as a stick-slip event, or the extension of pseudopods, as illustrated in Fig. 8. Although the sum of all the fast processes is ultimately the cause of the changes in $\theta$, we model this sum simply as a random number, independent of $\varphi$. That is, we model the orientation $\theta$, of the intrinsic vector as a random walk and ignore its causal connection with $\varphi$:

$$\frac{d\theta}{dt} = \sqrt{2D_\theta}\eta_\theta, \qquad (1)$$

where $D_\theta$ is a diffusion coefficient and $\eta_\theta(t)$ is a normalized, Gaussian, white noise term whose mean is zero.

The time-development of the instantaneous direction of motion is governed by

$$\varphi(t) = \theta(t) + y(t), \qquad (2)$$

where the second term on the right-hand-side $y(t)$ is colored noise: It is the solution to the second order stochastic differential equation describing a

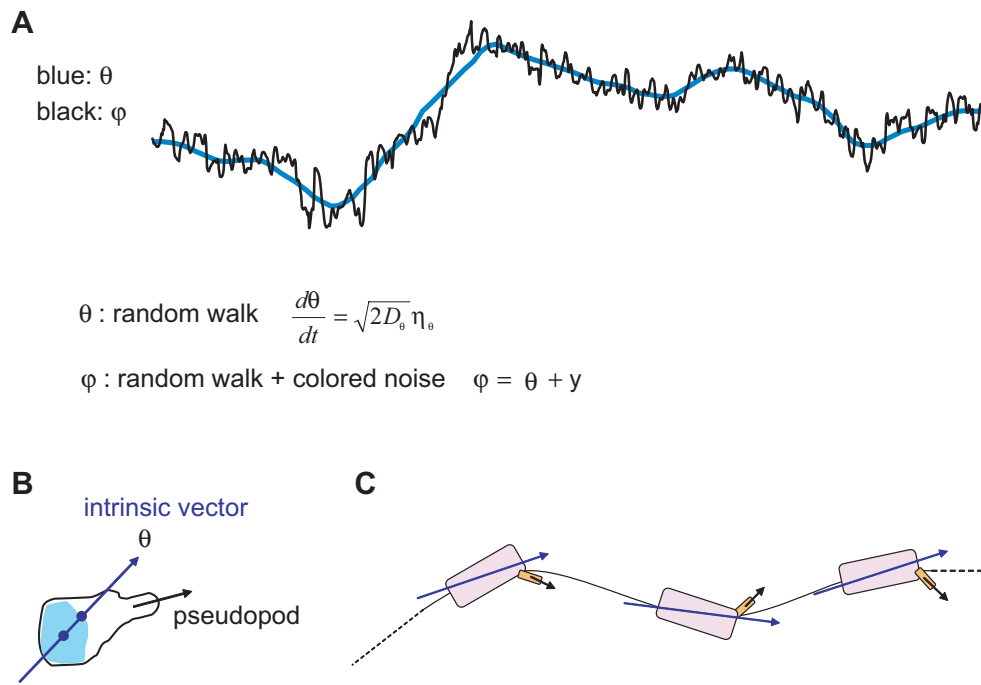

**Figure 8. A model for *Dictyostelium* motion.** (A) The blue line is hand drawn to guide the eye and represents $\theta$, a one-dimensional random walk. The black line $\varphi$, is an enlarged view of the data in box 3 of Fig. 6. In our model, $\varphi$ is the net effect of $\theta$ and colored noise centered on $\theta$. The stochastic differential equations used to describe the behavior of $\varphi$ and $\theta$ are explained in detail in the text. (B) $\theta$ is the angle assumed to be fixed by a cell's intrinsic polarity, possibly a vector directed from the center of nucleus (light blue) to the position of the centrosome [19]. Black arrow, pseudopod extensions and retractions lead to stochastic oscillations. New pseudopods bifurcate from old, and they swing back and forth about the internal vector. (C) A cartoon describing directional control.

noise-driven, harmonic oscillator with resonance at $2\pi f_0 = \sqrt{\kappa/m}$

$$m\frac{d^2 y(t)}{dt^2} = -\gamma \frac{dy(t)}{dt} - \kappa y(t) + \gamma\sqrt{2D_y}\eta_y(t), \quad (3)$$

where $m$ is a persistence parameter (in a mechanical model of a block on a spring it would be the inertial mass, but mass plays no role here), $\gamma$ is a dissipative parameter (friction in a mechanical model), $\kappa$ is a restoring-force parameter (spring constant), $D_y$ gives the strength of the driving force (noise) $\eta_y(t)$, which is normalized and has zero mean, but is not necessarily Gaussian, or white.

Does this model adequately capture the essential features of the data? We calculated the expected power spectral density (PSD) of $\varphi$ (Appendix), and compared the results to our experimental PSD (Fig. 9A). We also included the aliasing effect of finite sampling frequency and introduced a noise term accounting for measurement errors (dotted line; and see Appendix). Fig. 9A shows that our model fits the experimental PSD well. Two time scales were obtained from the fitting parameters and found to be consistent with other analyses discussed here. The value for $f_0$, the resonance frequency, gives a characteristic time of 2.4 ± 0.1 min, consistent with the average oscillation period observed in the time series of $\varphi$. $\theta$ is the overall direction of motion, and its value is subject to a random walk (Eq. 1). Given the fitted diffusion coefficient $D_\theta$, we estimate that after 1 rad²/$D_\theta$ ≈ 8 min, the MSD of $\theta$ grows large enough that we can consider cells to have lost their original direction. This is also consistent with the persistence time described in Fig. 3.

In order to further test our model we ran Monte-Carlo simulations of Eqs. 1-3 with parameters obtained from a fit to the experimental PSD. We then subjected the synthetic data to the same analysis as the experimental data. As an example, Fig. 9B shows the close agreement between the measured autocorrelation function of $\Delta\varphi(t)=\varphi(t)-\varphi(t-\tau)$ for both experimental and synthetic data. We also formed the histogram of $\Delta\varphi$s and found that it follows a Laplacian (double-exponential) distribution, Fig. 9C. The same distribution was found for the synthetic data (data not shown). A Laplacian distribution describes the difference between two independent, identically distributed, exponential, random variables, implying that $\varphi$ itself is exponentially distributed. Such exponential distributions for motility data are thought to arise from the interplay of various

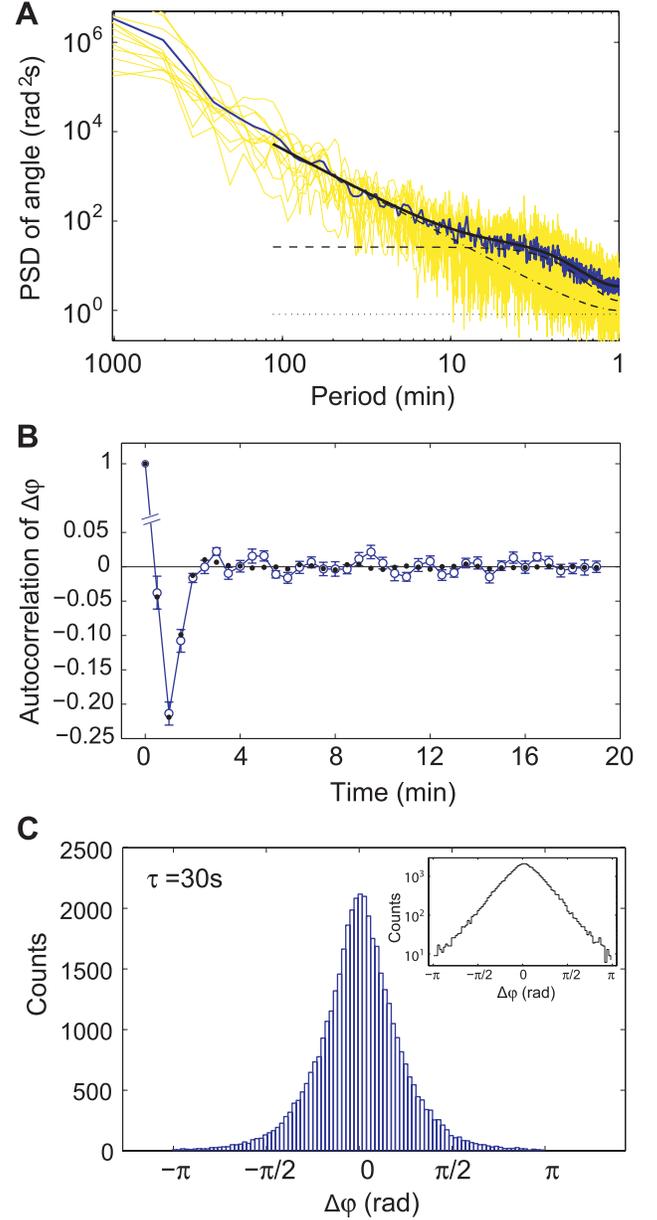

Figure 9. **Statistics of the cumulative angles and fit to the theory.** (**A**) Experimental power spectral density (PSD) of $\varphi$ and fit of the theory to the data. Two time-scales were returned by the fit: $(f_0)^{-1}$ = 2.35 ± 0.08 min and 1 rad² / $D_\theta$ = 7.6 ± 0.3 min, approximately the duration of a pseudopod and the time it takes for a cell to lose its sense of direction, respectively. Twelve individual PSDs, one for each cell, are shown in yellow. The average over the cells is shown in blue. The solid black line is a fit of the theory to the averaged signal. Dashed, dash-dotted, and dotted lines indicate the contribution to the PSD for the colored noise, the random walk, and the tracking-error terms, respectively (see Appendix for details). (**B**) Autocorrelation function for $\Delta\varphi$. $\varphi$s were calculated for $\tau$ = 30s. Blue: Experimental values and standard errors. Black: Theoretical expectation value calculated from a Monte Carlo simulation on $\varphi$ based on Eqs. 1-3, with parameters obtained from a fit to the PSD. (**C**) Experimental histogram of $\Delta\varphi$s calculated for $\tau$ = 30s. Insert: Same histogram shown on a semi-logarithm scale demonstrating the non-Gaussian, exponential tails. With increasing $\tau$, the distribution of $\Delta\varphi$ becomes more and more Gaussian (data not shown).

cellular processes combined with a finite rate of ATP production [20].

*Dictyostelium* cells move by continuously extending new pseudopods. A restoring force exists because pseudopods cannot extend to infinity, and thus oscillations should be restricted about θ. The driving force in our model is a sum of white noise contributed by all lateral pseudopods, and an oscillating force applied by the leading pseudopods. A colored noise term captures these contributions. Normally, a pseudopod leads cell motion for 1-2 min before a new pseudopod forms and dominates, consistent with the observed periods of oscillations shown in Fig. 6.

**A quantitative picture of cell motion**

The observation that cumulative angles oscillate, and the way that cells move by extending pseudopods, inspired us to describe cell motion as a discrete, but easily pictured turn-run-turn model.

The anti-correlations found in turn directions is consistent with the discovery that new leading pseudopods arise mostly by dividing old ones [21]. These correlations can be understood in a quantitative manner with the smallest possible set of assumptions. Next to having no memory at all (as in the flipping of a coin), the simplest kind of memory is one that only extends one step back in the past. Processes that have this kind of memory are called Markov processes. Assume that the turns are given by a Markov process and ascribe a value of +1 to left-turns and -1 to right-turns, or vice versa. The autocorrelation function for the turns is then easily found to be $c_m = (p - q)^m$, where $m$ is the time-lag measured in turns, $p$ is the probability of making a turn in the same direction (left-left or right-right) and $q$ is the probability of making a turn in the opposite direction (left-right or right-left). Both $p$ and $q$ are known to us, since $p + q = 1$ and $q/p = 2.1 \pm 0.1$ (see Results). The near-exact equivalence of the data and the theory shown in Fig. 7C supports this interpretation of the turn-memory in *Dictyostelium*.

Taken together, the data presented in Figs. 7 & 9 support the following descriptive model of migration: Each cell is polarized (has an intrinsic vector) and moves (runs) in a random direction for a random period of time by the extension of pseudopods. Because the duration of a run is exponentially distributed and uncorrelated with any previous run duration it is tempting to think of each run as corresponding to an actin-polymerization event nucleated by a protein complex anchored to the leading edge of a pseudopod. If the association (dissociation) of this actin-nucleating protein complex with the membrane is equally probable at all times, it would be described by a Poisson process and the time between on/off events would follow an exponential distribution. Why the direction of migration follows a double-exponential distribution is less obvious, but may reveal that there are just a few limiting steps underlying the behavior documented in Fig. 7D.

We may compare our results to the pioneering work of others with *Dictyostelium* and the results of Hartman *et al* with neutrophils [22,23]. Potel and Mackay carried out a very thorough analysis of aggregation-competent *Dictyostelium* cells plated at a cell density orders of magnitude larger than the one we have used. In their experiments, cell-cell interactions were common. Their results on cell speed as a function of time, and change in direction between successive times as an approximate demonstration of a persistence time, are in broad agreement with ours. However, their time resolution, and the fact that the cells were responding to each other's presence, confounded their results, and in the end they concluded that a simple, persistent, random walk best fit their data. Hartman *et al* found local non-Markov displacements in their study with neutrophils. Although their time resolution was high, their analyses were applied to cell traces that are typically only a few cell-diameters long, and thus their experiments did not last long enough to detect the features we report here.

**An efficient search strategy?**

In the Introduction we described three well-studied models have been used to characterize searching behavior (Fig. 1). *Dictyostelium* motion differs from each of them. Unlike the intermittent behavior in the two-state model, no obvious searching and relocating phases are observed. However, we note that *Dictyostelium* cells appear to have gone one step further by not pausing on the search, and detecting signals as they move. Instead of choosing turns randomly in amplitude and direction (random walk) or sampling from a power law distributed run length (a Lévy walk), both *Dictyostelium* and *Polysphondylium* bias their motion by remembering their last turn and employ a persistence time of ~9 min. This means that cells cover more territory in a given number of steps than they would in a random walk (Fig. 1A,D). Also, they pick turn amplitudes randomly from an exponential distribution rather than a Gaussian distribution, which has been suggested to help optimize search efficiency when targets are randomly distributed in a patch of finite size [24].

The character of this search algorithm was examined by comparing our results to straight-line and random walk searches with respect to the efficiency of searching, defined as the number of targets captured per unit time, the number of cell diameters over which a cell can recognize a close-by target, and the target density. In these simulations, which assume that targets are arrayed at random on an infinite plane, straight-line searching is the most efficient because for a given detection radius $R$ all deviations from a straight line cover some fraction of the area already covered; and, as expected, the efficiency is lowest for random walk searches. *Dictyostelium* cells are about half as efficient as a straight-line search, and 1.6 to 2.4 fold more efficient than random walk searching.

The improvement in search efficiency relative to a random walk is due to the fairly straight motion of the amoebae. Although the cells move by the extension of a series of discrete pseudopods, the direction of protrusion of these processes is coordinated in a manner that gives rise to an overall directed motion: Each new pseudopod propels the cell in a direction slightly angled relative to the general direction of motion, but the pseudopods are generated in a left-right-left-right fashion such that the cell zigzags it's way forward. This propensity to zigzag is quantitatively described by the anti-correlation between turns, see Fig. 7 A-C. Without the anti-correlation, the persistence length would drop from the observed value of 48 μm to just 20 μm (calculated from the equivalent two-dimensional freely-rotating-chain model, with segment length = 5 μm and angle = 0.67 radians).

In addition, the comparatively long persistence time of ~9 min we discovered might help amoebae not only while foraging on their own, but also during the earliest stages of multi-cellular life when starving cells begin to signal to each other. The signaling system consists of cyclic AMP waves that propagate as spirals or circles from a core of signaling cells. These waves continue to organize morphogenesis as cells begin to stream towards the signaling centers guided by the traveling wave front. During the earliest stages of signaling, wave periodicity is 6-10 min, depending on the strain and growth conditions. This time corresponds closely to the 9 min persistence time reported here, and suggests to us that once cells are given an orientation in the wave, they propagate towards the center without further information from the cyclic AMP gradient. This may also help explain how cells discriminate between the back of a receding wave and the front of an approaching one. They use the persistence time once oriented to move in a more or less straight line toward the center and away from the periphery. Because this bias is in our view stochastic in nature, this might also help explain why cells occasionally move in the wrong direction, *i.e.* away from the signaling center.

## Summary
We have discovered that *Dictyostelium* and *Polysphondylium* cell motion is not a simple random walk. Unlike a Lévy walk, no intrinsic scale invariance in cell trajectory is apparent. Unlike an Ornstein-Uhlenbeck process, cell velocity distributions deviate from a Gaussian velocity distribution. Unlike a worm-like-chain model, the observed oscillations in angles indicate a well-developed and organized cellular mechanism driving the observed behavior. With respect to searching strategy, a left turn tends to be followed by a right turn. Cells move forward in a zig-zag manner and maintain a long directional persistence. In this way, time wasted on exhaustive back and forth searching is greatly reduced, thereby enlarging the search area and improving search efficiency.

## MATERIALS and METHODS
### Cell Culture
*Dictyostelium discoideum* AX4 and *Polysphondylium pallidum* PN500 were grown on lawns of *Escherichia coli* B/r at 22°C as described [25]. Vegetative amoebae were harvested and bacteria removed by centrifugation. The cells were suspended in PB (20 mM $KH_2PO_4$, 20 mM $Na_2HPO_4 \cdot 7H_2O$) and plated at densities of ~1 cell/$cm^2$ on 2% agar in distilled deionized water. At this cell density the ratio of cell area / agar surface area is ~$4 \times 10^{-6}$.

### Cell Tracking
Cell movement was followed by phase contrast microscopy using a 10X objective. Movies were recorded at 10 sec intervals for 8 to 10 hr. (Movie S1, S2).

### Data Analysis
Cell locations were defined as the centroids of a cell's contours (Movie S1, S2). The trajectory of each centroid consisted of a sequence of paired coordinates $\Delta t$ =10 sec apart: $\vec{r}_j = \vec{r}(t_j), t_j = j\Delta t, j = 1,2,3...$. A displacement between any two cell positions was defined as: $\vec{s}_j(\tau) = \vec{s}_j(n\Delta t) = \vec{r}_j - \vec{r}_{j-n}, n = 1,2,3...$. Velocities were then calculated:

$$\vec{v}_j(\tau) = \vec{v}_j(n\Delta t) = \vec{s}_j(n\Delta t)/n\Delta t = \vec{s}_j(\tau)/\tau, n = 1,2,3...$$
. Instantaneous angles were calculated: $\beta_j = \operatorname{atan}[(y_j - y_{j-1})/(x_j - x_{j-1})]$, then corrected by +/- $2\pi$ if the changes between consecutive angles were larger than $\pi$, and added to the previous angle: $\varphi_j = \varphi_{j-1} + \beta_j$. Also, if the change in two consecutive angles was bigger than $\pi$, it was identified as a false jump. The jump was then replaced by two random numbers from a Gaussian distribution with zero mean and the same standard deviation as obtained from experimental data. Further analysis and original programming was carried out in MATLAB.

**Recognizing turns**
Cell trajectories were first smoothed by a moving average over 5 consecutive positions ($\vec{r}_j' = (\vec{r}_{j-2} + \vec{r}_{j-1} + \vec{r}_j + \vec{r}_{j+1} + \vec{r}_{j+2})/5$). Then the time series of angles ($\varphi_j$) and changes in consecutive angles ($\Delta\varphi_j = \varphi_j - \varphi_{j-1}$) were calculated with $\tau = 10s$ using smoothed positions. $\Delta\varphi_j$ represent cell turning rates as a function of time. When its amplitude goes above a threshold value, a cell is considered to be making a turn and the corresponding time points were marked. Next, the marked time points were clustered: First, all consecutive points were clustered and the largest cluster-member was picked to represent this turn; Second, if the time interval between any two consecutive turns was less than 30 sec, it was considered to be part of the same turn and again the largest value was picked. Each cell had an individual threshold value, chosen as the average amplitude of that cell's $\Delta\varphi$ series. If the turn had a positive value for $\Delta\varphi$ it was recorded as a left turn (counterclockwise), otherwise, as a right turn (clockwise).

**Cumulative angle analysis: PSD, autocorrelation and histogram**
To reduce noise in the analysis of cumulative angles (Fig. 9), $\varphi$ was calculated with $\tau = 30s$ using unprocessed original positions. Thus, from each cell trajectory, 3 interlaced time series of $\varphi$ were obtained and analyzed.

**Simulating Search Efficiency**
A Monte-Carlo simulation was employed to compare *Dictyostelium* trajectories to random walks and straight-line searches. Targets were distributed randomly on an infinite plane with a characteristic average density. Searchers using different strategies were compared: Amoeboid motions were parameterized using our experimental data, random walks were simulated using the average speed for *Dictyostelium*, and straight-line movement at a constant speed used the mean speed of these amoebae. If a target site lay within a 'detection radius' (R) from the searcher, then this target was scored as found and removed. Efficiency was defined as the ratio of the mean number of targets found to the total traveling time. In each simulation, cells were allowed to search for 10 hr. The simulation was repeated with a range of target densities ($10^{-2}\mu m^{-2}$ - $10^{-5}\mu m^{-2}$), and the detection radius was varied from 5μm to 75μm, 1/3 to 5 cell diameters, respectively.

**APPENDIX:**
**Simulation of cumulative angles**
We simulated the harmonic noise by first solving Eq. 3 on the discrete time-interval $\Delta t$ (= 30 sec) to obtain the recursive relation:
$$\begin{pmatrix} y_{j+1} \\ u_{j+1} \end{pmatrix} = e^{-M\Delta t} \begin{pmatrix} y_j \\ u_j \end{pmatrix} + \begin{pmatrix} \Delta y_j \\ \Delta u_j \end{pmatrix}, \quad M = \begin{bmatrix} 0 & -1 \\ \kappa/m & \gamma/m \end{bmatrix},$$
where $u = dy/dt$, and we demand that $\Delta y_j$ and $\Delta u_j$ be exponentially distributed random numbers of zero mean and flat power spectrum. Note that the distributions for $\Delta y_j$ and $\Delta u_j$ depend on $\Delta t$ because exponential distributions, unlike Gaussian ones, are not stable under integration. That is, as $\Delta t$ is increased, these distributions will tend to Gaussians.

**Power spectral analysis**
The frequency content of $\varphi$ was examined by Fourier-transforming the two independent equations of motion, Eqs. 1 & 3,
$$-2\pi i f_k \tilde{\theta}_k = \sqrt{2D_\theta}\tilde{\eta}_{\theta,k}$$
and
$$(-2\pi i f_k)^2 m\tilde{y}_k = 2\pi i f_k \gamma \tilde{y}_k - \kappa \tilde{y}_k + \sqrt{2D_y}\gamma\tilde{\eta}_{y,k}$$
where
$$\tilde{x}_k = \int_{-t_{msr}/2}^{t_{msr}/2} x(t) e^{i2\pi f_k t} dt$$
$x = \theta, y, \eta_\theta$ or $\eta_\varphi$, and $t_{msr}$ is the measurement time. From which we form the power spectral density:
$$P_k \equiv \frac{\langle |\tilde{\varphi}_k|^2 \rangle}{t_{msr}} = \frac{D_\phi/(2\pi^2)}{f_k^2} + \frac{D_y/(2\pi^2)}{(\frac{2\pi m}{\gamma})^2(f_0^2 - f_k^2)^2 + f_k^2}$$
Where we have used the following characteristics of the noise terms
$$\langle \tilde{\eta}_{\theta,k} \rangle = \langle \tilde{\eta}_{y,k} \rangle = 0; \quad \langle \tilde{\eta}_{\theta,k}^* \tilde{\eta}_{\theta,l} \rangle = \langle \tilde{\eta}_{y,k}^* \tilde{\eta}_{y,l} \rangle = t_{msr}\delta_{k,l};$$

$$\langle \tilde{\eta}_{\theta,k} \tilde{\eta}_{y,l} \rangle = 0;$$

We see that the PSD consist of two terms, one that decays as $f^{-2}$ (first term on the right-hand-side corresponding to free diffusion) and one that, potentially, has a resonance peak at $f_k=f_0$ (second term on the right-hand-side).

**Aliasing/finite sampling frequency**

For reasons of mathematical ease, the above treatment implicitly assumed continuous sampling. In an experiment, however, data are taken at discrete time-intervals, leading to aliasing. This means that, for frequencies near $f_{sample}$, the measured power can be more than 100% larger than predicted from the above theory [26]. Taking aliasing into account is straightforward and introduces no extra fit-parameters, as shown below.

The aliased version of the first term, corresponding to free diffusion, is:

$$P_k^{(alias)} = \frac{D_\theta / f_{sample}^2}{1 - \cos(2\pi f_k / f_{sample})}$$

The correct aliased version of the second term is considerably more involved and we refer the reader to (Nørrelykke & Flyvbjerg, unpublished). For completeness, we also include a term for the measurement error in the fit. The measurement error is assumed to be white noise and thus has a flat power spectrum, so its shape is unaltered by aliasing.

We zero-padded time-traces of $\varphi$ to the nearest power of two greater than the longest trace, before fast Fourier transforming. Zero-padding artificially increases the frequency resolution of the PSD. This results in the PSD values no longer being independent of each other, so the error-bars in the residual plot underestimate the true standard error.

**Least squares fitting**

We fitted the above expression to the experimental PSD in the least-squares sense (via the built-in lsqnlin procedure in MatLab), using as weights $P_k^{(alias)}$. That is, we minimized:

$$\chi^2 = \sum \left( \frac{P_k^{(alias)} - P_k^{(ex)}}{P_k^{(alias)} / \sqrt{n}} \right)^2$$

Least squares fitting presupposes each data point to be drawn from a Gaussian distribution – which is not the case here! Rather, the power at a given frequency, averaged over $n$ individual PSDs, is Gamma-distributed and tends to a Gaussian distribution as $n \to \infty$ (Nørrelykke & Flyvbjerg, unpublished).

A Gamma-distribution is skewed, and this skewness leads to an overestimate of some of the fit-parameters by a factor of $1/n$ when doing least-squares fitting (Nørrelykke & Flyvbjerg, unpublished). But this effect is well understood, and was taken into account.

The fit parameters obtained were: $f_0$=0.0072 rad$^{-1}$s$^{-1}$ ± 3.5% ; $\gamma/m$=0.067 s$^{-1}$ ± 9.4% ; $D_\gamma$=0.012 rad$^2$s$^{-1}$ ± 3.2% ; and $D_\theta$=0.023 rad$^2$s$^{-1}$ ± 4.3% . A harmonic oscillator has three qualitatively different solutions, depending on whether the parameter $\zeta = \gamma/\sqrt{4\kappa m}$ is greater than, equal to, or smaller than unity. In the first case ($\zeta > 1$) the system in over-damped and no oscillations occur; in the third ($\zeta < 1$) the system is under-damped and displays exponentially decaying oscillations; when $\zeta = 1$ the system is poised at a critical point separating the two classes of behavior. In our case $\zeta = 0.74 \pm$ 12.9%, that is, we are in the under-damped, oscillating regime, but still rather close to the critical point, which is why we observe no clear resonance peak in the power spectrum.

The error-bars cited on the fit parameters were taken from the formal covariance matrix, calculated as the inverse of the Jacobian matrix of $\chi^2$ multiplied by its own complex conjugate. Since the residuals are neither independent nor Gaussian distributed these variances are only estimates of the true ones.

**Acknowledgments**

We thank S. Schvartsman, I. Golding, H. Flyvbjerg, R. Austin, S. Sawai, J. Shaevitz, and F. Jülicher for discussions, and all members of the Cox laboratory for their help and advice. This paper benefited greatly from the comments of an anonymous referee.

# Supplementary Information

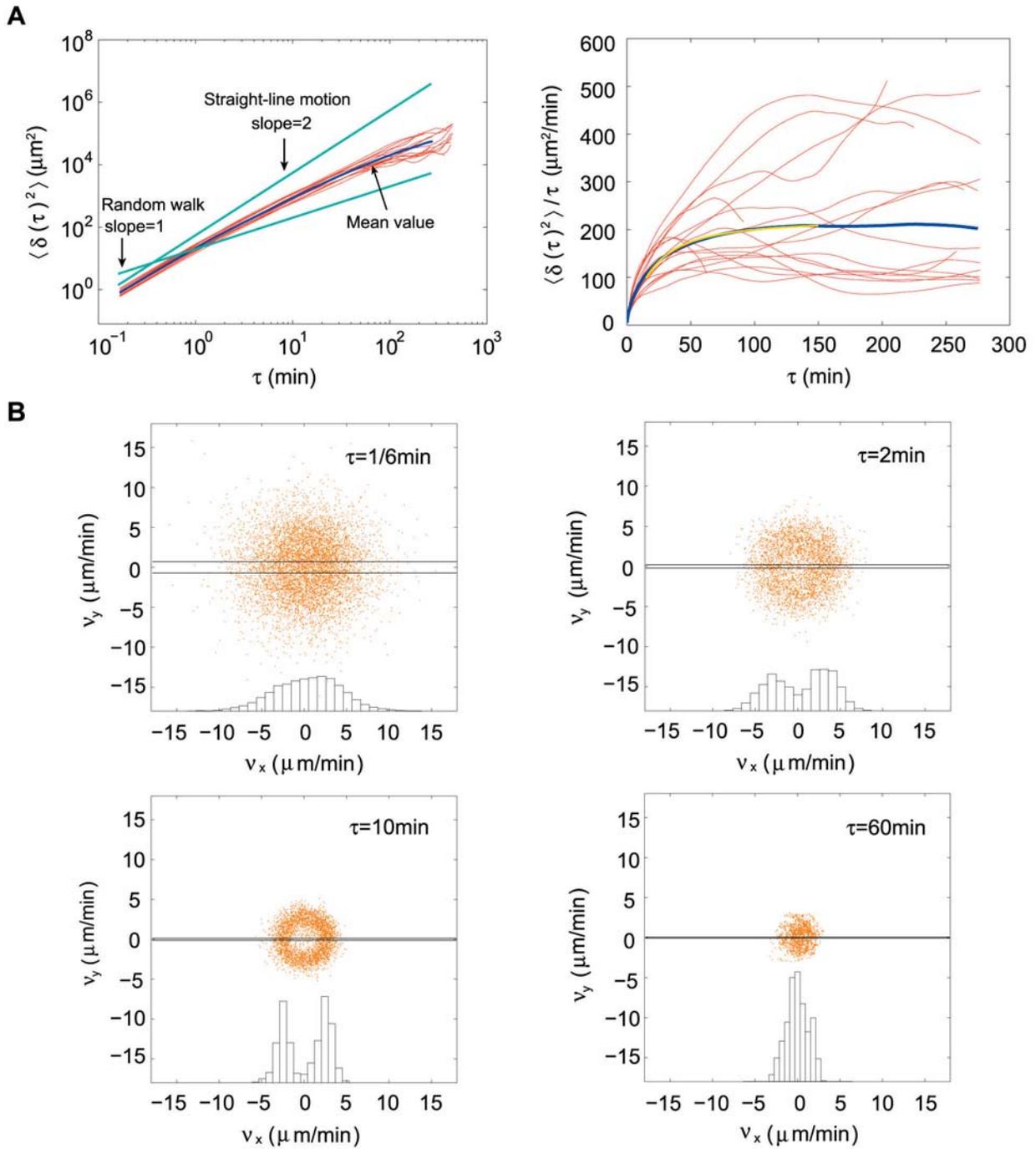

**Figure S1.** *Polysphondylium* **motion on an agar surface.** (**A**) Mean-squared displacement plotted for 17 cells. Yellow, fit of an exponential cross-over from directed to random motion in the interval $\tau$ [15:150] min: $\langle \delta(\tau)^2 \rangle = 2 t_p v^2 (\tau - t_p(1 - \exp(-\tau/t_p)))$, where $v = 3.1 \pm 0.1$ μm/min is a characteristic speed, and $t_p = 11.7 \pm 0.2$ min is a persistence time. (**B**) Cell velocity distributions are non-Gaussian.

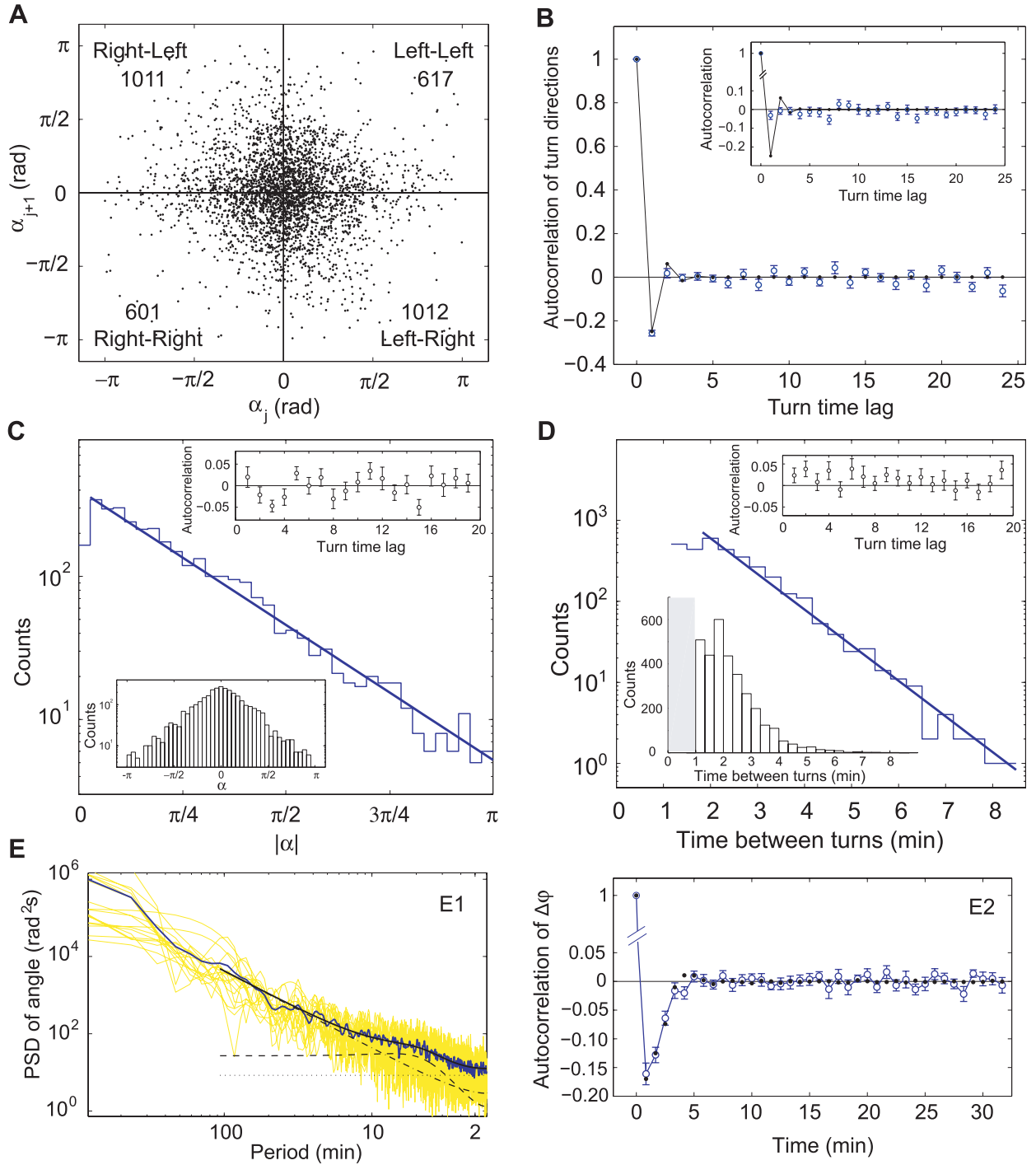

**Figure S2.** *Polysphondylium* **motion on an agar surface.** (**A**) The $j^{th}$ turn plotted against the $(j+1)^{th}$ turn for the data from all 17 trajectories. There are 2023 data points in the second and fourth quadrant, 1218 in the first and third, and thus the $(j+1)^{th}$ turn is biased by the $j^{th}$ turn by a factor of 1.7. (**B**) Autocorrelation function for the turn directions. Blue: Experimental values and standard errors. Black: Theoretical expectation value for a Markov process with probabilities taken from panel A. Insert: Verification that turn-correlations are real and not an artifact of the turn-detection algorithm. Blue: Autocorrelation function for synthetic data. The angle-dynamics was simulated by a worm-like-chain model (WLC) with parameters taken from the MSD of the real data. A small, negative, artifactual correlation is detected which extends for around 3 turns. Black: Same as the main-panel, shown for comparison. (**C**) Histogram of turn amplitudes. These data are well fitted by an exponential distribution (characteristic angle = 0.72 rad). Lower left panel: Histogram of $\alpha$. Upper right panel: Autocorrelation function for turn amplitudes, no correlation was observed. (**D**) Histogram of time intervals between detected turns. These data are well fitted by an exponential distribution (characteristic time = 0.98 min). Data is from all 17 trajectories. Lower left panel: Same histogram but on linear scale. The smallest detected value for $t_{j+1}-t_j$ is 1 min, the cut-off shown by the grey bar. Upper right panel: Normalized autocorrelation function for time between turns. No significant correlations were observed, consistent with a Poisson process. (**E1**) Experimental power spectral density of $\varphi$ is well fitted by theory. Two time-scales were returned by the fit: $(f_0)^{-1} = 4.3 \pm 0.2$ min and $1\ \mathrm{rad}^2 / D_\theta = 7.9 \pm 0.5$ min. (**E2**) Experimental autocorrelation function for $\Delta\varphi$ (blue) is consistent with the theoretical expectation (black) calculated from a simulation. $\varphi$s were calculated for $\tau = 50$s in E1 and E2.

15